\documentclass[runningheads]{llncs}
\usepackage[T1]{fontenc}
\usepackage[nolist]{acronym}
\usepackage{enumerate}
\usepackage{graphicx}
\usepackage{enumitem}
\usepackage{circledsteps}
\usepackage{listings}
\usepackage{subcaption}
\usepackage{hyperref}
\usepackage{array}
\usepackage{multirow}
\usepackage[english]{babel}
\usepackage{listings}
\usepackage{booktabs}
\usepackage[pages=some]{background}
\begin{acronym}
    \acro{afl}[AFL]{American Fuzzy Lop}
    \acro{soc}[SoC]{System-On-Chip}
    \acro{put}[PUT]{Program Under Test}
    \acro{mmio}[MMIO]{Memory-Mapped I/O}
    \acro{vcml}[VCML]{Virtual Components Modelling Library}
    \acro{iss}[ISS]{Instruction Set Simulator}
    \acro{fss}[FSS]{Full-System Simulator}
    \acro{vp}[VP]{Virtual Prototype}
    \acro{tlm}[TLM]{Transaction-Level Modeling}
    \acro{dbt}[DBT]{Dynamic Binary Translation}
    \acro{avp}[AVP]{Arm Virtual Platform}
    \acro{nvic}[NVIC]{Nested Vector Interrupt Controller}
    \acro{ftl}[FTL]{Fast Translation Library}
    \acro{os}[OS]{Operating System}
    \acro{mcu}[MCU]{Microcontroller}
    \acro{adc}[ADC]{Analog-to-Digital Converter}
    \acro{rtc}[RTC]{Real-Time Clock}
\end{acronym}

\usepackage{colortbl} 
\usepackage{tikz} 
\usepackage{pgfplots} 
\usepackage{xcolor} 
\definecolor{rwth_blue}{RGB}{0,84,159}
\usepackage{tcolorbox} 
\tcbuselibrary{listings}
\newcommand\myCircled[2][]{\ifmmode
\Circled[fill color=white,inner color=black,#1]{\mathsf{#2}}
\else
\Circled[fill color=white,inner color=black,#1]{\sffamily#2}
\fi
}
\newcommand{\copyrighttext}{
  \footnotesize \textcopyright~2025 
  Chiara Ghinami et al. This is a preprint accepted at the 
  \textit{25th International Conference on Embedded Computer Systems: Architectures, Modeling and Simulation (SAMOS XXV)}. 
  The final version will be published by Springer.
}

\definecolor{dkgreen}{rgb}{0,0.6,0}
\definecolor{gray}{rgb}{0.5,0.5,0.5}
\definecolor{mauve}{rgb}{0.58,0,0.82}

\begin{document}

\renewcommand{\thelstlisting}{\arabic{lstlisting}}
\setcounter{lstlisting}{0}
\title{Leveraging SystemC-TLM-based Virtual Prototypes for Embedded Software Fuzzing}
\titlerunning{Leveraging SystemC-based VPs for Embedded Fuzzing}
%
\author{Chiara Ghinami\inst{1} \and
Jonas Winzer\inst{1} \and
Nils Bosbach\inst{1} \and
Lennart M. Reimann\inst{1} \and
Lukas Jünger\inst{2} \and
Simon Wörner\inst{3} \and
Rainer Leupers\inst{1}}
\authorrunning{C. Ghinami et al.}
%
\institute{RWTH Aachen University, Aachen Germany \and
MachineWare GmbH, Aachen Germany \and
CISPA Helmholtz Center for Information Security, Saarbrücken, Germany}

\newcommand\copyrightnotice{%
    \backgroundsetup{opacity=1, scale=1, angle=0, contents={
            \color{black}%
            \begin{tikzpicture}[remember picture,overlay]%
                \node[anchor=south,yshift=10pt] at (current page.south) {\fbox{\parbox{\dimexpr0.75\textwidth-\fboxsep-\fboxrule\relax}{\copyrighttext}}};
                \node[anchor=north,yshift=-10pt,text=gray] at (current page.north) {\shortstack[c]{\large PREPRINT - accepted by the \textit{25th International Conference on Embedded Computer Systems:}\\\textit{Architectures, Modeling and Simulation (SAMOS XXV)}}};
            \end{tikzpicture}%
        }%
    }%
    \BgThispage%
}

\maketitle
\copyrightnotice
\begin{abstract}
SystemC-based virtual prototypes have emerged as widely adopted tools to test software ahead of hardware availability, reducing the time-to-market and improving software reliability. 
Recently, fuzzing has become a popular method for automated software testing due to its ability to quickly identify corner-case errors. However, its application to embedded software is still limited. Simulator tools can help bridge this gap by providing a more powerful and controlled execution environment for testing. Existing solutions, however, often tightly couple fuzzers with built-in simulators that lack support for hardware peripherals and offer limited flexibility, restricting their ability to test embedded software. To address these limitations, we present a framework that allows the integration of American-Fuzzy-Lop-based fuzzers and SystemC-based simulators. The framework provides a harness to decouple the adopted fuzzer and simulator. In addition, it intercepts peripheral accesses and queries the fuzzer for values, effectively linking peripheral behavior to the fuzzer. This solution enables flexible interchangeability of peripherals within the simulation environment and supports the interfacing of different SystemC-based virtual prototypes. The flexibility of the proposed solution is demonstrated by integrating the harness with different simulators and by testing various softwares.

\keywords{Virtual Prototypes \and SystemC \and Fuzzing \and AFL.}
\end{abstract}

\section{Introduction}
In embedded system development, where time-to-market and reliability are critical, \acp{vp} have become essential. They enable early software development and debugging in virtual environments well before physical hardware is available. The SystemC standard~\cite{systemc2025}, with its \ac{tlm} extension, is the \textit{de-facto} standard for \ac{soc} simulation, supporting flexible and scalable modeling of heterogeneous systems.

Among software testing techniques, greybox fuzzing has emerged as a powerful technique that repeatedly executes the \ac{put} with varying inputs to detect faults autonomously. \ac{afl}\cite{zalewski2016american}, now extended as \ac{afl}++\cite{fioraldi2020afl++}, integrates many advanced fuzzing strategies~\cite{yun2022fuzzing,fioraldi2023dissecting}. Given its versatility and popularity, \ac{afl}++ was chosen for this study.

While effective for general-purpose software, fuzzing embedded systems is more complex. Direct testing on hardware suffers from scalability and performance constraints~\cite{scharnowski2022fuzzware,yun2022fuzzing}. Firmware rehosting addresses this by executing the software in simulators~\cite{chen2016towards,fasano2021sok,feng2020p2im}. However, QEMU-based solutions~\cite{10.5555/1247360.1247401} are often tightly coupled with fuzzers, lack extensibility~\cite{fioraldi2022libafl}, and do not support peripheral, limiting their applicability to complex hardware-dependent systems.

To address these limitations, we present a framework\footnote{Available at \url{https://github.com/Jonaswinz/AFLplusplus}} that connects \ac{afl}-based fuzzers with any SystemC-based \ac{vp}. A custom harness bridges the fuzzer and simulator, enabling integration without modifying the fuzzer or duplicating fuzzing logic within the simulator. We demonstrate its flexibility by interfacing it with two different SystemC-based simulators.

Additionally, we introduce a plug-in for \ac{mmio} tracking in the simulator that intercepts peripheral reads and retrieves values from the fuzzer. This enables efficient peripheral fuzzing with minimal integration effort. We validate our approach on both a bare-metal application and a Zephyr~\cite{zephyr} \ac{os}-based system.

\section{Background}
In this section, we give an overview of various simulator technologies and the SystemC standard. Then, we introduce fuzz testing and discuss the difficulties of applying fuzzing to embedded programming. 

\subsection{Virtual Prototypes} \label{systemc_vps}
SystemC~\cite{systemc2025} is a C++-based framework for hardware modeling, supporting both low- and high-level abstractions. This work focuses on \ac{tlm}, which enables standardized, transaction-level communication between components, abstracting low-level hardware details.

QEMU~\cite{10.5555/1247360.1247401} is a widely used system simulator employing \ac{dbt}, but lacks native SystemC support, limiting extensibility. To bridge this gap, the open-source \ac{avp} simulator~\cite{10.1145/3300189.3300191} wraps QEMU in a SystemC environment, enabling \ac{tlm}-based simulation for ARM Cortex-A/M systems. Similarly, MachineWare's proprietary SIM-A and SIM-V simulators~\cite{Machineware,simv} use SystemC together with the \ac{ftl} for high-speed execution.

This study targets Cortex-M0, using SIM-A and the 32-bit AVP variant (AVP32). Both rely on the \ac{vcml} library~\cite{vcml}, which provides modular peripheral models. New peripherals, interconnects, and any custom logic needed for a specific application can be defined and added to the simulator.

\begin{figure}[t!]
\centerline{\includegraphics[totalheight=5cm]{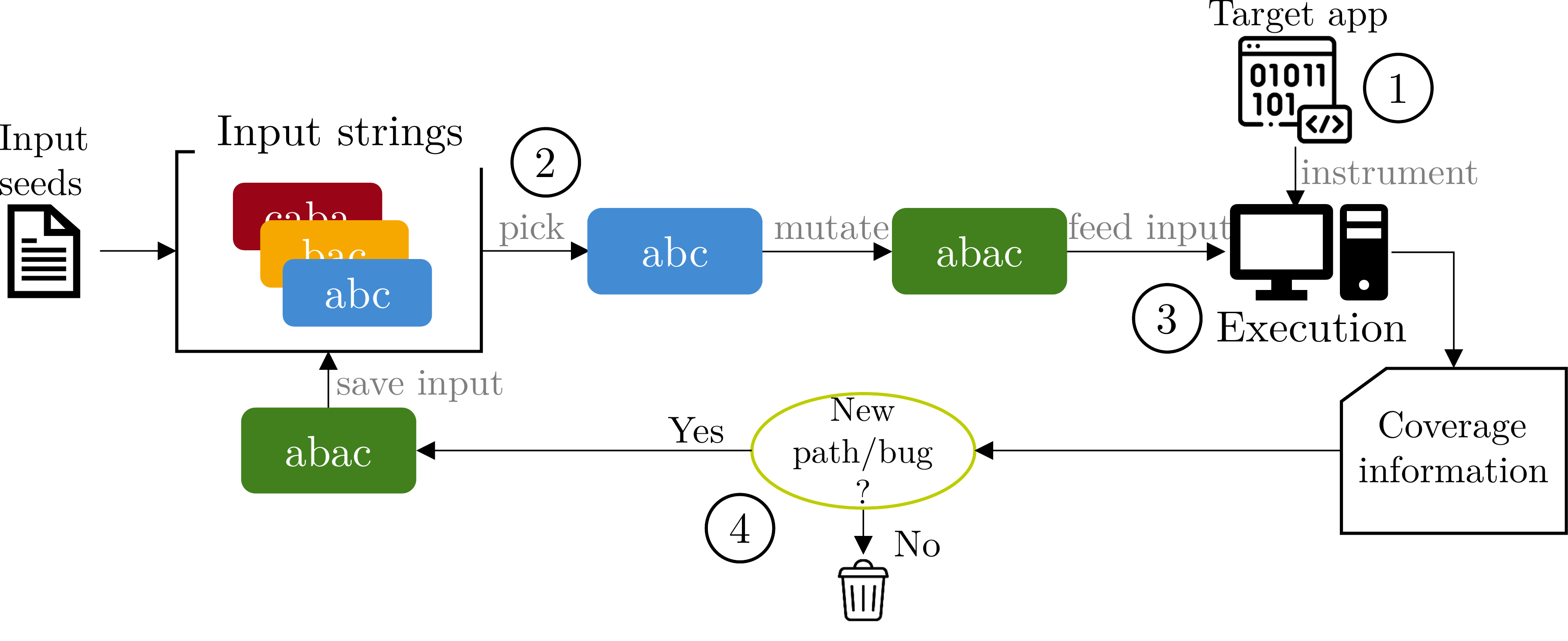}}
    \caption{The fuzz testing loop.}
    \label{fig:fuzzing}
\end{figure}

\subsection{Fuzz Testing} \label{fuzz_testing}
Fuzzing is a widely used technique for assessing software correctness and reliability by automatically generating and executing test inputs. Greybox fuzzers like \ac{afl} leverage coverage feedback to guide input mutation and explore new execution paths (Fig.~\ref{fig:fuzzing}). They typically start with an initial seed input, which may be provided by the user. When source code is available, \ac{afl} instruments the program to collect coverage data; otherwise, it uses a binary translator (e.g., user-mode QEMU~\cite{10.5555/1247360.1247401}) to execute the binary and extract coverage. However, QEMU lacks peripheral support, making it unsuitable for embedded software.

\section{Related Work}
Recent research has explored embedded fuzzing, with Fuzzware~\cite{scharnowski2022fuzzware} being most closely related to our work. It uses QEMU-AFL to fuzz firmware by tracking \ac{mmio} accesses via QEMU callbacks, avoiding peripheral modeling. To tackle hard-to-reach states, it integrates a symbolic execution engine to guide input generation. While this reduces manual modeling effort, it introduces overhead and state explosion, limiting scalability for complex software.

Firmadyne~\cite{chen2016towards} uses full-system QEMU with a custom kernel to support firmware emulation, though it does not involve fuzzing. Similarly, FirmAFL~\cite{10.5555/3361338.3361415} combines user- and full-system QEMU modes to balance speed and fidelity. However, both approaches inherit QEMU’s limitations in extensibility and modularity, which SystemC-based \acp{vp} address more effectively.

\section{Contribution} 
In Section~\ref{mmio_tracking}, we explain the implementation of MMIO tracking. Then, in Section~\ref{framework}, we describe the proposed fuzzing framework and how it exploits MMIO tracking. Our objective is to test an application that interacts with peripherals, using a fuzzer to generate test cases for the \ac{put}. 

\begin{figure}[t!]
\centerline{\includegraphics[totalheight=4cm]{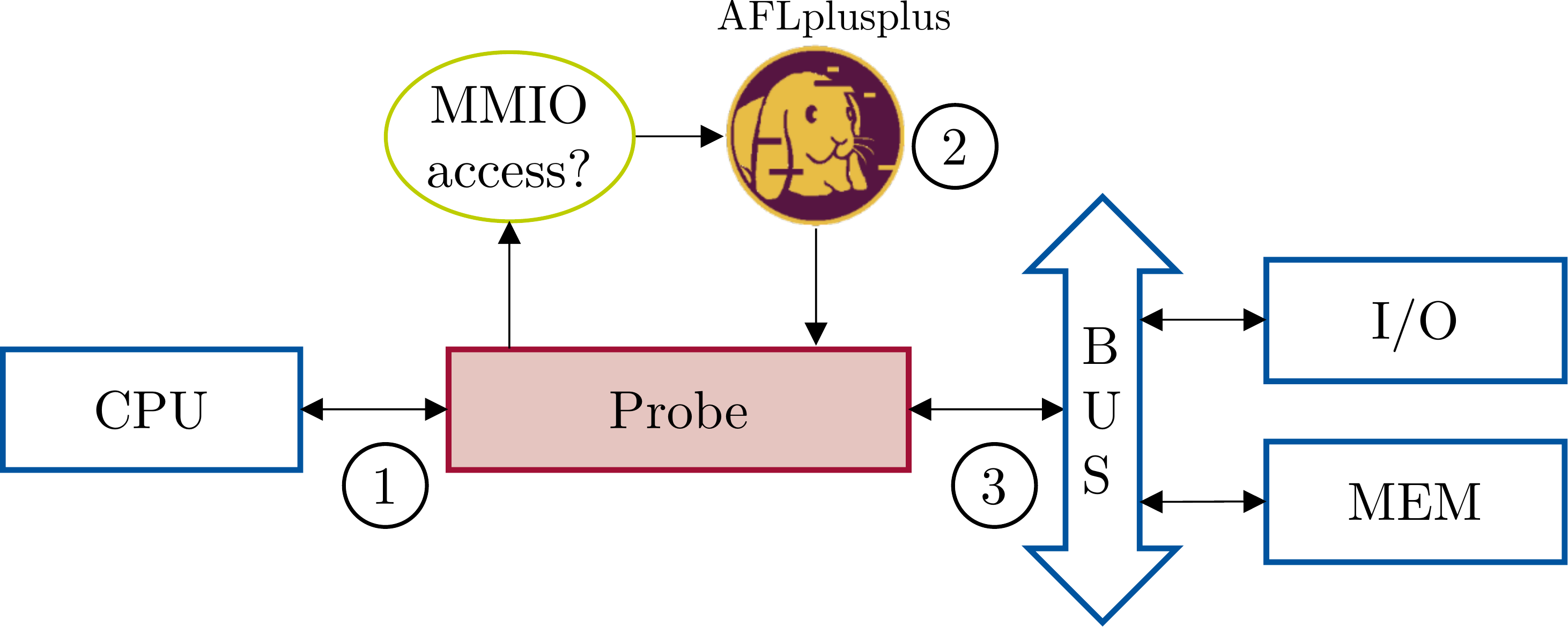}}
    \caption{The \ac{vp} with the probe component.}
    \label{fig:mmio_tracking}
\end{figure}

\subsection{MMIO Tracking Plug-in} \label{mmio_tracking}
Whenever the program performs a read operation within the peripheral’s address range, the tracking system intercepts the request and retrieves the value generated by the fuzzer (Fig.~\ref{fig:mmio_tracking}).
We implemented MMIO tracking by adding a plug-in, called \textit{probe} to the \ac{vp} which sits between the CPU and the bus, and intercepts the \ac{tlm} transactions to the peripherals. The user can specify the address range to be considered. During the program execution, the probe intercepts the read access request from the CPU \myCircled{1}. If the address matches the one specified by the user, it reads the fuzzer value \myCircled{2} and returns it to the CPU \myCircled{3}.

\subsection{Proposed Fuzzing Framework} \label{framework}
The AFL fuzzer supports running various user-mode simulators, such as user-mode QEMU, Unicorn, and others. Our goal is to integrate the fuzzer with a SystemC-TLM-based full-system simulator, while minimizing the need to modify the fuzzer for each new simulator. To achieve this, we introduce a new option that allows the AFL fuzzer to interact with a harness that then manages the communication with the simulator.

\subsubsection{VP configuration} \label{vp_initialization}
The simulator can run standalone or in fuzzing mode with a few additional features enabled. MMIO tracking must be activated, with target addresses defined in a configuration file. The VP also collects code coverage, and crash detection is implemented via breakpoints:

\begin{itemize}
\item \textbf{Bare-metal applications without error handler}: Following the ARM 32-bit calling convention, when a function returns a single value, the return value is placed in the R0 register. To monitor if an error occurred, we tracked the returned value from the main routine by reading the value of the R0 register. If the value is 1, it indicates that an error occurred.
\item \textbf{OSs and bare-metal applications with an error handler}: When dealing with programs that have an error handler function, we set a breakpoint to the error handler that, if executed, signals that an error occurred.
\end{itemize}

\subsubsection{Execution}
After the \ac{vp} configuration, the simulation starts. When an MMIO read access to the tracked peripheral occurs, the \ac{vp} reads a fuzzer's input. If an error occurs, the VP notifies the harness, which in turn sends the information to the fuzzer and restarts the \ac{vp} process. If the simulation ends without errors, the \ac{vp} simply notifies the harness that the simulation is over and sends the coverage information to the fuzzer. To reduce the overhead of restarting the VP for every test run, we introduce the \textit{persistent mode}. We added the possibility of setting an entry and exit address of the program, once the exit address is hit, the \ac{vp} jumps to the entry point, thus reducing the overhead of restarting the VP for every test run.

\section{Results}

To validate the proposed framework, we tested the application shown in Listing~\ref{lst:uart_baremetal}. In this program, we brute-force a password received via UART and check the read string. When the received string is equal to the \textit{password} string, the program ends with an error. This is a typical application for challenging the ability of a fuzzer to spot corner cases \cite{scharnowski2022fuzzware}.
To increase the computational complexity of the program and thus to obtain more realistic comparisons, we included the Caesar Cipher algorithm in the received string. It is a simple encryption algorithm that replaces every letter by shifting it by a fixed number of positions in the alphabet.

We simulated a commercial ARM Cortex-M0-based SoCs, the nRF51~\cite{nrf51} from Nordic Semiconductor, we tracked the UART read accesses, and we included the timer peripheral model in the simulation without tracking it. Among the many peripherals excluded from the simulated environment are the GPIOs, most of the timers (with only one included), as well as SPI, I2C, CAN, and others. We performed all the tests on a 12-core Intel(R) Core(TM) i7-1255U CPU. In Section~\ref{execution_modes}, we compare the two modes of execution while in Section~\ref{qemu_comparison} we compare our tool with QEMU-AFL. 

\begin{lstlisting}[language=C, caption=UART baremetal example, label={lst:uart_baremetal}]
int i=0; char read_c; char read_str[128]; 
do{
    read_c = uart_receive();
    read_str[i++] = read_c;
} while(read_c!='\n' && read_c!='\0');
read_str[i-1] = 0;

caesar_cipher(read_str, 1);

if(!strcmp(encr_password, read_str)) 
    exit(1); //error

exit(0);

\end{lstlisting}

\subsection{Execution Modes Comparison} \label{execution_modes}
In this Section, we compare the persistent mode with the basic set-up of our framework that simply restarts the VP for each execution. We run a bare-metal version of the password example and integrated it in a Zephyr \ac{os} application. We run the experiments using the AVP32 and SIM-A simulators introduced in Section~\ref{systemc_vps}. 

\begin{table}[b!]
\centering
\caption{Execution speed (in milliseconds) of different VP stages: VP startup, VP configuration and program execution.\vspace{0.2cm}}
\begin{tabular}{p{2cm} p{1.5cm} p{1.5cm} p{1.5cm} p{1.5cm} p{1.5cm} p{1.5cm}}
\multirow{2}{*}{\textbf{Software}} & \multicolumn{2}{c}{\textbf{Startup}} & \multicolumn{2}{c}{\textbf{Configuration}} & \multicolumn{2}{c}{\textbf{Execution}} \\ [0.5ex]
                            & AVP & SIM-A & AVP & SIM-A & AVP & SIM-A \\ 
\hline \hline
\rowcolor{black!15}\textbf{Bare Metal}         & 21 & 120 & 0.2 & 0.2 & 0.9 & 0.4 \\

\textbf{Zephyr}             & 30 & 130 & 0.2 & 0.2 & 25.0 & 15.0 \\ [0.5ex]
\end{tabular}

\label{table:2}
\end{table}


Table~\ref{table:2} provides the measurements for different stages. The startup time refers to the time needed to create the VP process and for the VP to be ready to execute the program. During the configuration stage, breakpoints are set, and MMIO tracking is enabled. In the execution phase, the VP runs the \ac{put}. In the Zephyr example, the execution time includes the initialization of the peripherals and the time to boot. As shown in Table~\ref{table:2}, the VP startup time is the bottleneck. This latency is higher for the SIM-A simulator than for the AVP because of the usage of a license server at startup time to check out the software license. Even though the Zephyr example executes the same application as the bare metal, the execution time is higher because of the underlying OS.

\begin{figure}[t]
    \centering
    \ref*{mylegend}

    \begin{subfigure}[b]{0.48\textwidth}
\begin{tikzpicture}[scale=1]
    \begin{axis}[
        ybar,
        enlargelimits=false, 
        scaled y ticks = false,
        legend style={at={(3,6.2)},    
                    anchor=north,legend columns=4},         bar width=0.5cm,
        width=6cm,
        height=4.3cm,
        ylabel={execs/second},
        ymax=4900,
        ymin=-0.050,
        enlarge x limits=0.4,
        xticklabels={AVP,SIM-A},
        xtick=data,
        legend to name={mylegend},
        ylabel near ticks,
        ylabel shift={-0.55em},
        xtick pos=left,
        ytick pos=left,
        legend image post style={scale=0.9},
        y tick label style={font=\small},
        y label style={font=\small},
        every node near coord/.append style={
                        anchor=west,
                        rotate=90
                },
    ]
    \addplot[fill=rwth_blue!25, nodes near coords] coordinates {(0,35) (1,7)}; 
     \addplot[fill=rwth_blue!75, nodes near coords] coordinates {(0,1950) (1,3100) }; 

    \legend{Restart, Persistent}
    \end{axis}
    \end{tikzpicture}  
    
    \caption{\label{fig:baremetal_bar}Baremetal.}      
    \end{subfigure}
        \hfill
    \begin{subfigure}[b]{0.48\textwidth}
\begin{tikzpicture}[scale=1]
    \begin{axis}[
        ybar,
        enlargelimits=false, 
        scaled y ticks = false,
        legend style={at={(0.495,1.23)}, anchor=north,legend columns=-1},
        bar width=0.5cm,
        width=6cm,
        height=4.3cm,
        ylabel={execs/second},
        ymax=4900,
        ymin=-0.050,
        enlarge x limits=0.4,
        xticklabels={AVP,SIM-A},
        xtick=data,
        ylabel near ticks,
        ylabel shift={-0.55em},
        xtick pos=left,
        ytick pos=left,
        legend image post style={scale=0.9},
        y tick label style={font=\small},
        y label style={font=\small},
        every node near coord/.append style={
                        anchor=west,
                        rotate=90
                },
    ]
    \addplot[fill=rwth_blue!25, nodes near coords] coordinates {(0,20) (1,6.5)}; 
     \addplot[fill=rwth_blue!75, nodes near coords] coordinates {(0,1800) (1,2800) }; 

    \end{axis}
    \end{tikzpicture}  
    
    \caption{\label{fig:zephyr_bar}Zephyr OS.}      
    \end{subfigure}
    \caption{\label{fig:baremetal_zephyr_res}The simulator's performance for different execution modes.}
\end{figure}
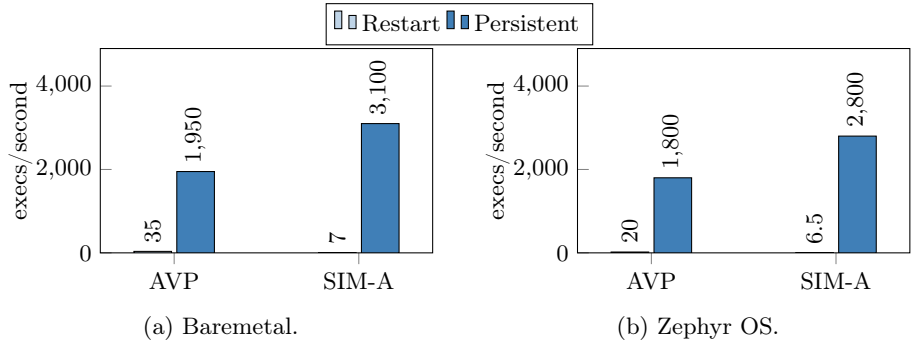

In Figure~\ref{fig:baremetal_zephyr_res} the execution modes are compared. As a consequence of the high VP startup time, the restart mode has the significantly worst execution-per-second value. As shown in the figure, the persistent mode is particularly beneficial for the Zephyr example, since it avoids the need of rebooting the OS at each execution. 

\begin{figure}[t]
    \centering
\begin{tikzpicture}[scale=1]
    \begin{axis}[
        ybar,
        enlargelimits=false, 
        scaled y ticks = false,
        legend style={at={(0.495,1.23)}, anchor=north,legend columns=-1},
        bar width=0.3cm,
        width=6cm,
        height=4.3cm,
        ylabel={execs/second},
        ymax=4900,
        ymin=-0.050,
        enlarge x limits=0.13,
        symbolic x coords={AVP,SIM-A,QEMU},
        ylabel near ticks,
        ylabel shift={-0.55em},
        xtick pos=left,
        ytick pos=left,
        legend image post style={scale=0.9},
        y tick label style={font=\small},
        y label style={font=\small},
        every node near coord/.append style={
                        anchor=west,
                        rotate=90
                },
    ]
    \addplot[fill=rwth_blue!75, nodes near coords] coordinates {(AVP,1950) (SIM-A,3100) (QEMU,3300)};
     \addplot[fill=rwth_blue!25, nodes near coords] coordinates {(AVP,1800) (SIM-A,2800) };

    \legend{Bare metal, Zephyr OS}
    \end{axis}
    \end{tikzpicture}  
    
    \caption{\label{fig:sim_comparison}Comparison of AVP32, SIM-A and QEMU executing in persistent mode.}      
\end{figure}
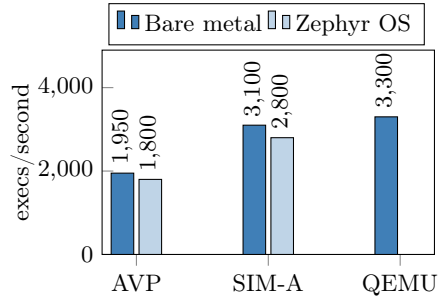

\subsection{QEMU-AFL Comparison} \label{qemu_comparison}
In this Section, we compare our framework with QEMU-AFL. Unlike SIM-A and AVP, QEMU-AFL operates in user mode and lacks access to peripherals, which allows it to achieve higher execution speeds, but also prevents the execution of Zephyr. As shown in Fig.~\ref{fig:sim_comparison}, a high-performance simulator such as SIM-A can bridge the gap with QEMU-AFL by utilizing the Fast Transition Level Library (FTL) that guarantees high simulation speed.

\section{Conclusion}
In this work, we focused on fuzzing embedded systems using SystemC-based VPs. We successfully interfaced multiple simulators with the AFL fuzzer by developing a flexible framework that manages VP instances and facilitates communication with the fuzzer. This approach maintains a clear separation between the fuzzer and the VP tool, allowing for their independent usage, testing and development. Our framework is compatible with any SystemC-based VP, leveraging the SystemC standardized interface, which also enables full customizability of simulators to support and simulate various hardware platforms. To ensure compatibility with the AFL++ fuzzer, we instrumented the VPs, including a plug-in for tracking MMIO accesses, enabling users to specify which peripheral interactions to monitor. The system was deployed on two different simulators and used to test a bare-metal and an OS application. The framework demonstrated to reach comparable performance to state-of-the-art QEMU-AFL while integrating additional peripheral support, which allows to test a wider range of embedded applications. Future work will expand its application to test and debug a broader range of Zephyr drivers and Arduino libraries, targeting more complex SoCs, such as multi-core heterogeneous SoCs.


\bibliographystyle{plainurl}
\bibliography{bibtexentry}
\end{document}